\documentclass[review]{elsarticle} %review  or  3p

\usepackage{hyperref}
\usepackage{changepage}
\journal{Engineering Fracture Mechanics}

\usepackage{lmodern}
\usepackage{lineno}

\usepackage{amsmath} %added for maths environments (equation and align)
\usepackage{amssymb}
\usepackage{upgreek}
\usepackage{mathtools, nccmath} %added for \xrightleftharpoons
\biboptions{sort&compress}
\usepackage{makecell}  
\usepackage{float} % To use H in floating figures
\usepackage{tablefootnote}
\usepackage{subcaption} %allowing for subcaptions and subfigures
\usepackage[margin=2cm]{geometry}% TO PLAY WITH THE MARGINS
\usepackage[capitalise, nameinlink]{cleveref}

\usepackage{bm} %bold symbols by using \bm

\usepackage{placeins}

\renewcommand{\v}[1]{\bm{\mathrm{#1}}}
\newcommand{\m}[1]{\boldsymbol{#1}} 

\begin{document}

\begin{frontmatter}
\title{Cohesive phase-field fracture with an explicit strength surface: an eigenstrain-based return-mapping formulation}

\author[1]{Tim Hageman\corref{A1}}
\cortext[A1]{Corresponding author}
\ead{tim.hageman@eng.ox.ac.uk}

\address[1]{Department of Engineering Science, University of Oxford, Oxford OX1 3PJ, UK}

\begin{abstract} 
    Standard phase-field fracture methods are rooted in brittle fracture theory and therefore do not inherently prescribe a material strength for crack nucleation, while also struggling to capture cohesive fracture behaviour. Recent eigenstrain-based formulations overcome both limitations by introducing fracture eigenstrains that decouple the strength surface from the fracture energy, but their implementation has so far relied on direct energy-minimization frameworks rather than standard finite-element procedures. In this work, we exploit the fact that the eigenstrains require no spatial gradients and reformulate the eigenstrain evolution as a local constitutive model, analogous to those used in plasticity, that is resolved at each integration point. As a result, the cohesive phase-field requires no additional global degrees of freedom beyond those of a standard phase-field formulation and can be readily integrated into existing finite-element codes. Two strength criteria are considered: a non-smooth criterion with independent tensile and shear strengths, and a smooth Drucker-Prager-like criterion that captures pressure-dependent strengthening under compression. Consistent tangent operators are derived for both criteria, ensuring robust convergence of the global Newton-Raphson solver. The framework is validated against three benchmark problems: a plate with a hole under tension and compression, a single-edge notched plate under shear, and a notched plate under dynamic loading. The results demonstrate mesh-independent and phase-field length-scale-independent behaviour, confirm that the fracture energy governs the transition between brittle and cohesive regimes, and show that complex phenomena such as crack branching under dynamic loading are naturally captured. All source codes are openly available.
\end{abstract}

% 3-5 Highlights (max 85 char per highlight)
% \begin{highlights}
% \item Cohesive phase-field fracture with an explicit, tuneable strength surface.
% \item Eigenstrains are updated by local return mapping at integration points.
% \item Formulation is implemented in a standard finite-element framework.
% \item Benchmark cases show mesh-independence and phase-field length scale independence.
% \item Code used for all benchmark cases is openly available online.
% \end{highlights}


\end{frontmatter}
%\linenumbers

\section*{Nomenclature}
\begin{tabular}{@{}ll@{\qquad\qquad}ll@{}}
$\v{b}$ & Body forces & $c$ & Damping coefficient \\
$d(\phi)$ & Degradation function & $\m{D}_\text{tan}$ & Consistent tangent stiffness \\
$E$ & Young's modulus & $f_\text{s}$ & Shear strength \\
$f_\text{t}$ & Tensile strength & $F_\text{d}$ & Degradable strength potential \\
$F_\text{i}$ & Non-degradable strength potential & $\m{G}$, $\m{G}_i$ & Eigenstrain directions \\
$G_\text{c}$ & Fracture energy release rate & $\m{I}$ & Identity matrix \\
$K$ & Bulk modulus & $\ell$ & Phase-field length scale \\
$\m{L}$ & Strain--displacement operator & $\m{N}_\text{u}$, $\v{N}_\phi$ & Shape functions \\
$\v{u}$ & Displacement & $\gamma(\phi)$ & Phase-field distribution function \\
$\m{\varepsilon}$ & Total strain & $\m{\varepsilon}_\text{el}$ & Elastic strain \\
$\varepsilon_\text{ref}$ & Reference strain (DP criterion) & $\m{\eta}$ & Fracture eigenstrain \\
$\kappa$ & Residual strength parameter & $\kappa_\text{t}$ & Residual stiffness parameter \\
$\lambda$, $\lambda_i$ & Eigenstrain multipliers & $\mu$ & Shear modulus \\
$\nu$ & Poisson's ratio & $\rho$ & Density \\
$\m{\sigma}$ & Cauchy stress & $\v{\tau}$ & External traction \\
$\phi$ & Phase-field parameter & $\Psi$ & Total energy \\
$\Omega$ & Domain & $\Gamma_\text{c}$ & Crack surface \\
$\mathcal{F}$ & Crack driving force (history) & $\text{tr}(\cdot)$ & Trace operator \\
$\text{dev}(\cdot)$ & Deviatoric operator & $\lVert\cdot\rVert$ & $L_2$ norm \\
$\dot{(\cdot)}$, $\ddot{(\cdot)}$ & First, second time derivative & $\left<\cdot\right>^{\pm}$ & Macaulay brackets \\
\end{tabular}

\section{Introduction}
\label{sec:intro}

Over the last decades, the phase-field method has become one of the most popular methods to model fracture propagation \citep{Egger2019a, deBorst2022}. Fractures are described through a partial-differential equation rather than by explicitly inserting new fracture faces, making integration within existing finite element software straightforward \citep{Navidtehrani2021, Chen2022a, Molnar2017}. As a result, this method allows for arbitrary fracture paths, crack branching, and merging without special considerations of these features \citep{Bourdin2014, Hofacker2013}. Phase-field models have also been combined with a range of other physical phenomena, e.g., the numerous works on hydraulic fractures \citep{Wheeler2014, Clayton2025, Sun2021, Miehe2015b} and chemical degradation \citep{Wu2016, Duda2018, Hageman2023b, Mandal2021}, where treating cracks as a diffuse interface makes implementing these additional phenomena straightforward. 

The phase-field fracture method is based on Griffith's theory, prescribing the energy dissipated by propagating cracks and letting the fracture path itself follow from minimizing the total energy contained within the system. However, this origin in brittle fracture mechanics has also imposed a strong limitation \citep{Kristensen2021, Lopez-Pamies2025}: Griffith's theory does not capture when cracks nucleate as it does not prescribe a material strength. Instead, it solely predicts when cracks propagate and the energy dissipated during this propagation \citep{Griffith1921}. As a result, phase-field methods inherited the same issues, with crack propagation being modelled in a thermodynamically consistent manner, while crack nucleation often uses ad-hoc criteria. For instance, while the phase-field length scale is introduced as a regularization parameter, defining the distance over which the discrete crack is distributed, treating it as a material parameter instead allows a material strength to be defined \citep{Tanne2018, Mandal2019}. Other approaches to capture this material strength that needs to be exceeded for cracks to nucleate include a series of stress-splitting schemes that split the elastic energy between a part that does, and does not, contribute to damage, thereby approximating strength surfaces \citep{Kumar2020, DeLorenzis2021, Vicentini2024, Zhang2022, Infante-Garcia2024}. 

A second limitation of standard phase-field methods is the difficulty to capture ductile behaviour, stemming again from the method's basis in brittle fracture mechanics. While the degradation function can be calibrated to match the unloading behaviour expected \citep{Geelen2019, Gupta2024}, this calibration is done through uni-axial tension cases, and extending this behaviour to multi-axial loading is often not possible \citep{Feng2023}. Furthermore, the damage functions do not include any fracture energy. As a result, calibration becomes material-specific, limiting application of the same models to a range of materials. Similarly, cohesive behaviour can be approximated based on the choice of phase-field distribution function \citep{Pham2011}, but this turns the choice of how to distribute the fracture energy over a finite region into a material modelling choice rather than a mathematical concept. While this limitation can be overcome by explicitly adding cohesive forces based on the displacement discontinuity, this adds extra complexity, requiring the displacement jump to be defined throughout the domain \citep{Verhoosel2013, GhaffariMotlagh2020}. Alternatively, the phase-field method can be combined with a plasticity model, using part of the plastic energy dissipated as driving force for fractures \citep{Miehe2016c, Ambati2015c, Alessi2018a}. However, if plasticity is only expected to occur near the crack tip, this approach is both computationally expensive, requiring a sufficiently fine mesh to resolve the plastic zone around the crack tip, and requires numerous additional constitutive modelling choices.

Recent work by Vicentini et al. \citep{Vicentini2026} and Bourdin et al. \citep{Bourdin2025} has started to overcome these limitations, explicitly introducing tuneable failure surfaces into phase-field formulations. By introducing fracture eigenstrains that decouple the fracture release energy from the material strength, and including an explicit threshold below which cracks cannot nucleate, their formulations are able to capture material behaviour across a range of fracture energies and material strengths without relying on tuning regularization parameters or damage functions. However, while these works provide a thorough mathematical basis, their implementations solve for the eigenstrains as additional field variables through global energy-minimization frameworks that rely on the symbolic differentiation and optimization capabilities of the programming libraries used. This introduces additional global degrees of freedom, requires specialized software, and makes integration into standard finite-element codes, where constitutive models are evaluated pointwise at integration points, non-trivial. As a result, these formulations have so far only been demonstrated on simple academic configurations.

The aim of this work is to overcome this implementation barrier by exploiting a key property of the eigenstrain formulation: the fracture eigenstrains require no spatial gradients and thus their evolution can be resolved entirely at the integration-point level, analogous to plasticity return-mapping schemes, without introducing any additional global degrees of freedom. However, realising this requires: (i) defining eigenstrain directions for each strength criterion, including a new Drucker-Prager-like criterion that captures pressure-dependent strengthening under compression, (ii) deriving consistent tangent operators in closed form for both smooth and non-smooth strength surfaces, handling the non-trivial switching between active and inactive failure modes, and (iii) addressing the numerical conditioning challenges that arise from the near-singular tangent matrices at the onset of fracture. The resulting framework can be used as a drop-in replacement for standard phase-field constitutive models in any existing finite-element code, without modifications to the global solver, element formulations, or solution algorithms. The energy expressions and governing equations are described in \cref{sec:gov_eq}, followed by a detailed description of the return-mapping scheme and its consistent tangent stiffness matrix in \cref{sec:implementation}. The computational framework is then benchmarked against a series of standard cases in \cref{sec:results}, specifically focussed on showing mesh-independent and phase-field length-scale-independent behaviour, the ability to capture both brittle and cohesive fracture regimes, and complex phenomena such as mixed-mode crack propagation and dynamic crack branching. All codes used for these benchmark cases are openly available, allowing for building upon or using the models for other applications, see the data availability section at the end of this paper.

\section{Governing equations} \label{sec:gov_eq}
We consider a domain $\Omega$, described through the displacement $\v{u}$ and the phase-field parameter $\phi$. $\phi$ indicates the loss of cohesive strength with $\phi=0$ indicating a fully intact material and $\phi=1$ indicating a full loss of strength. This is in contrast to standard phase-field descriptions where it is used to indicate the presence of cracks through a loss of the stiffness. Following the method from \citep{Vicentini2026}, the elastic strains are composed of a displacement component and a fracture-eigenstrain component, using a small-strains assumption:
\begin{equation} \label{eq:strain}
    \m{\varepsilon}_\text{el} = \m{\varepsilon} - \m{\eta} = \nabla^s \v{u} - \m{\eta}
\end{equation}
where the eigenstrains $\m{\eta}$ allow for the representation of crack-like behaviour, by reducing the elastic strains to represent the reduced stresses transferred through crack surfaces. This $\m{\eta}$ is zero when no cracks are present, is a small factor of $\m{\varepsilon}$ at the onset of cracks where the crack still allows for the transfer of stress through the cohesive zone, and evolves to a comparable magnitude as the total strains once the crack opening exceeds the cohesive length. This eigenstrain is also used to enforce different behaviour normal and tangential to the crack surface, e.g. allowing for normal stresses to represent self-contact under compression while removing the transfer of shear stresses. 

Throughout this work, we assume a linear-elastic bulk behaviour, such that the elastic energy stored in the material is given as a function of the elastic strains, \cref{eq:strain}, as:
\begin{equation} \label{eq:E_el}
    \Psi_\text{el} = \int_\Omega \frac{1}{2}K \text{tr}^2(\m{\varepsilon}-\m{\eta}) + \mu \lVert\text{dev}(\m{\varepsilon}-\m{\eta})\rVert^2\;\text{d}\Omega
\end{equation}
using the bulk modulus $K$ and the shear modulus $\mu$. $\text{tr}$ indicates the trace operator, while $\text{dev}()$ is used to indicate that only the deviatoric component of the strains is used, $\text{dev}(\m{\varepsilon}) = \m{\varepsilon} - \frac{1}{3}\text{tr}(\m{\varepsilon})\m{I}$. $\lVert \varepsilon \rVert$ indicates the $L_2$ norm, $\lVert \m{\varepsilon} \rVert^2=\m{\varepsilon}:\m{\varepsilon}$. We note that the use of a linear-elastic, small-strain model is not a requirement of the presented framework, and other constitutive models can equally be used for the bulk behaviour in a similar manner as standard phase-field models (e.g. finite strain \citep{Miehe2017a, Ambati2016a, Aldakheel2018} or hyperelasticity models \citep{Ang2022, Peng2020, Zhang2025, Mandal2020}). The assumption of linear elasticity is made here for simplicity and to focus on the implementation of the cohesive phase-field fracture model.

The fracture energy release rate, $G_\text{c}$, indicates the rate at which energy is released during fracture propagation, with low values of $G_\text{c}$ corresponding to brittle behaviour while higher values typically lead to a more ductile/cohesive type of fracture. Using this fracture energy, the energy dissipated through the opening of fractures is defined as \citep{Bourdin2000,Miehe2010}:
\begin{equation} \label{eq:E_Gc}
    \Psi_\text{c} = \int_{\Gamma_\text{c}} G_\text{c}\;\text{d}\Gamma_\text{c} \approx \int_\Omega \gamma(\phi) G_\text{c} \;\text{d}\Omega \qquad \frac{\partial \gamma(\phi) G_\text{c}}{\partial t}\geq 0
\end{equation}
with the KKT irreversibility condition ensuring that this is a purely dissipative mechanism \citep{Gerasimov2019}. The phase-field distribution function $\gamma(\phi)$ is used to distribute the energy released by a discrete crack over a finite region surrounding the crack. Specifically, a quadratic, AT2-type \citep{Ambrosio1990}, phase-field distribution function is used, given by:
\begin{equation} \label{eq:pf_Gamma}
    \gamma(\phi) = \frac{1}{2\ell}\left( \phi^2 + \ell^2 |\nabla \phi|^2\right)
\end{equation}
using the length scale $\ell$. This length scale indicates the region over which the loss of cohesive strength is distributed. Through the introduction of this length scale, mesh-size independent results can be attained even though the eigenstrains representing the crack are localized. We note that the choice to use an AT-2 model was made to not impose an additional strength criterion on top of the strength potentials defined in \cref{sec:strength}. Instead, using an AT-1 model would require the eigenstrains to exceed a certain length-scale dependent threshold before the phase-field variable can increase, thus resulting in initial cracks not starting to lose cohesion until this threshold is reached. As the eigenstrains are used to represent the crack opening, this would result in a non-physical behaviour where cracks can open without losing cohesion until a certain crack opening is reached. An AT-2 model does not have this issue, as the phase-field variable can increase as soon as the eigenstrains start to develop, allowing for a more physical representation of the cohesive behaviour.

To link the phase-field functions and the fracture eigenstrains, a material strength potential is introduced \citep{Vicentini2026}. As will be shown later, the derivative of this potential will define the maximum stress that can be sustained by the material, and how this degrades as the phase-field parameter increases. The energy contribution of this strength potential is given by:
\begin{equation}  \label{eq:E_F}
    \Psi_\text{f} = \int_\Omega F(\m{\eta})\;\text{d}\Omega = \int_\Omega d(\phi) F_\text{d}(\m{\eta})+F_\text{i}(\m{\eta}) \;\text{d}\Omega
\end{equation}
with the potentials $F_\text{d}$ and $F_\text{i}$ used in \cref{sec:strength} to define the strength of the material and the non-penetration condition respectively. The degradation function is defined in a standard manner:
\begin{equation} \label{eq:d_e}
    d(\phi) = (1-\kappa)(1-\phi)^2+\kappa
\end{equation}
such that the full strength potential $F_\text{d}$ is used to describe the intact material, while degrading this potential based on the phase-field parameter to represent the loss of cohesion as the cohesive crack evolves. The parameter $\kappa$ is used to provide a residual strength, preventing the strength potentials from degrading fully such that their derivatives remain well-defined. We note that this differs from standard phase-field formulations, where $\kappa$ is a residual stiffness. 

Finally, we include dynamic effects through inertia and damping energy contributions, and loading through the body and external forces energy contributions:
\begin{equation} \label{eq:E_other}
    \Psi_\text{i} = \int_\Omega \frac{1}{2}\rho \v{\dot{u}}^2 \;\text{d}\Omega + \int_t \int_\Omega \frac{1}{2}\rho c \v{\dot{u}}^2\;\text{d}t\text{d}\Omega
\qquad \qquad
    \Psi_\text{b} = \int_\Omega \v{b}^\top\v{u}\;\text{d}\Omega + \int_\Gamma \v{\tau}^\top \v{u}\;\text{d}\Gamma
\end{equation}
using the density $\rho$, body forces $\v{b}$, and external traction $\v{\tau}$. Within the subsequent examples, the damping is included to limit the impact of stress waves due to sudden material failure when considering cases under slow loading, whereas for cases explicitly considering dynamic loading it is set to zero. The inclusion of inertia, even for slow loading, helps with spreading the crack propagation over multiple time increments, preventing the need to resolve cracks propagating through the full domain within a single time/load step. Body forces are included within this derivation for completeness, but are set to $\v{b}=\v{0}$ in the benchmark cases considered. 

\subsection{Energy minimization}
The governing equations follow from minimizing the energies contained within the material, \cref{eq:E_el,eq:E_Gc,eq:E_F,eq:E_other}:
\begin{equation}
    \text{min}\left(\Psi(\v{u},\m{\eta},\phi)\right) = \text{min}( \Psi_\text{el} + \Psi_\text{c} + \Psi_\text{f} - \Psi_\text{i} - \Psi_\text{b}) 
\end{equation}
This is the total energy expression proposed by Vicentini et al. \citep{Vicentini2026}, who minimize it directly using a global optimization framework that treats $\v{u}$, $\m{\eta}$, and $\phi$ as independent field variables. In the present work, we instead derive the governing equations and constitutive relations that follow from this energy, and resolve the eigenstrain evolution locally at each integration point. The governing equations are given through the derivatives with regards to the displacement $\v{u}$ and phase-field parameter $\phi$. First, using the derivative with respect to $\v{u}$, we obtain the weak form of the momentum balance:
\begin{equation} \label{eq:momentum}
    0 = \frac{\partial \Psi}{\partial \v{u} } = \int_\Omega \delta\v{u}^\top\left(-\nabla\cdot \m{\sigma} + \rho c \v{\dot{u}} + \rho \v{\ddot{u}} - \v{b} \right)\;\text{d}\Omega 
\end{equation}
where integration by parts has been used to obtain the divergence of stress term, to more clearly convey the strong form relations. When producing the finite element forms in \cref{sec:implementation}, this will be reversed to remove the need to obtain stress gradients. Next, using the derivative of the total energy with respect to $\phi$, we obtain the weak form of the phase-field evolution:
\begin{equation} \label{eq:pf_evolution}
    0 = \frac{\partial \Psi}{\partial \phi} = \int_\Omega \delta \phi \left( \frac{G_\text{c}}{\ell}(\phi - \ell^2 \nabla^2 \phi) + \frac{\partial d(\phi)}{\partial \phi} \mathcal{F} \right)\;\text{d}\Omega
\end{equation}
Within this relation, the irreversibility condition is enforced by using the history parameter $\mathcal{F}$, defined as the maximum achieved crack driving force:
\begin{equation}
    \mathcal{F}^{t+\Delta t}(\m{\eta}) = \text{max}(F^{t+\Delta t}_\text{d}(\m{\eta}), \mathcal{F}^t)
\end{equation}
where the superscripts $t+\Delta t$ and $t$ are used to indicate the current and previous discrete time steps respectively. This history parameter ensures the crack formation is a purely dissipative mechanism, ensuring that the phase-field parameter $\phi$ can only increase, and thus the strength potential can only degrade, as the crack evolves.

In addition to these governing equations, the constitutive models result from minimizing the total energy with regard to our internal variables, the total strains and fracture eigenstrains. Considering the total strains results in the constitutive models for the elastic stresses:
\begin{equation} \label{eq:stresses}
    \m{\sigma} = \frac{\partial \Psi}{\partial \m{\varepsilon}} = K \text{tr}(\m{\varepsilon}-\m{\eta})\m{I} + 2\mu \text{dev}(\m{\varepsilon} - \m{\eta})
\end{equation}
and considering the fracture-eigenstrain results in the material strength criterion:
\begin{equation} \label{eq:strength_full}
    \bm{0} \leq \frac{\partial \Psi}{\partial \m{\eta}} = -\left(K \text{tr}(\m{\varepsilon}-\m{\eta})\m{I} + 2\mu \text{dev}(\m{\varepsilon} - \m{\eta})\right) + d(\phi) \frac{\partial F_\text{d}(\m{\eta})}{\partial \m{\eta}} + \frac{\partial F_\text{i}(\m{\eta})}{\partial \m{\eta}} = d(\phi) \frac{\partial F_\text{d}(\m{\eta})}{\partial \m{\eta}}+\frac{\partial F_\text{i}(\m{\eta})}{\partial \m{\eta}} - \m{\sigma}
\end{equation}
This constitutive model requires that the stress state of intact material remains within or on the derivative of the strength potential functions, $d(\phi) \frac{\partial F_\text{d}(\m{\eta})}{\partial \m{\eta}}+\frac{\partial F_\text{i}(\m{\eta})}{\partial \m{\eta}} \geq \m{\sigma}$. When the stress reaches this potential surface, the eigenstrains develop to retain the stress on the potential surface, while increasing the phase-field driving force $\mathcal{F}$. This then allows the phase-field variable $\phi$ to increase, decreasing the damage function $d$ such that the strength potential reduces. As such, the strength criterion decreases, representing the loss of cohesion as the eigenstrain (and thus the crack opening) increases. 

For the direction of the eigenstrains, we define these to follow the directions of the strains $\m{\varepsilon}$ through a function $\m{G}$, such that:
\begin{equation}
    \m{\eta} = \lambda \m{G}(\m{\varepsilon})
\end{equation}
in the case of a smooth failure surface, or in the case of a surface composed of multiple facets, using:
\begin{equation}
    \m{\eta} = \sum_i \lambda_i \m{G}_i(\m{\varepsilon})
\end{equation}
using the multipliers $\lambda$ to indicate the magnitude of the eigenstrains. Using these directions based on the strains $\m{\varepsilon}$ allows the eigenstrains $\m{\eta}$ to follow the same direction, thereby ensuring that the eigenstrains always act to reduce the total strains. As a result, we can reduce the strength criterion from \cref{eq:strength_full} to solely require the stresses in the direction of the currently applied strains not to exceed the fracture surface. This results in:
\begin{equation} \label{eq:StrengthCriteria}
    0\leq \frac{\partial \Psi}{\partial \lambda_i} = \frac{\partial \m{\eta}}{\partial \lambda_i} : \frac{\partial \Psi}{\partial \m{\eta}} = \left(\frac{\partial \m{\eta}}{\partial \lambda_i}\right) : \left( d(\phi) \frac{\partial F_\text{d}(\m{\eta})}{\partial \m{\eta}}+\frac{\partial F_\text{i}(\m{\eta})}{\partial \m{\eta}} - \m{\sigma} \right)
\end{equation}
This criterion only requires the stresses to be within the failure surface in the potential directions of unloading, thereby solely requiring the multipliers $\lambda$ to be resolved rather than all components of $\m{\eta}$. It should also be noted that, as no spatial gradients of the eigenstrains are required, these multipliers can be solved on a point-by-point basis, rather than requiring $\lambda$ to be resolved as a complete field. As a result, the constitutive models can be resolved in a similar manner as for plasticity schemes, using an integration-point-local return mapping scheme.

\subsection{Strength potentials} \label{sec:strength}

\begin{figure}
    \centering
    \includegraphics[clip=true,trim=0 15 0 15]{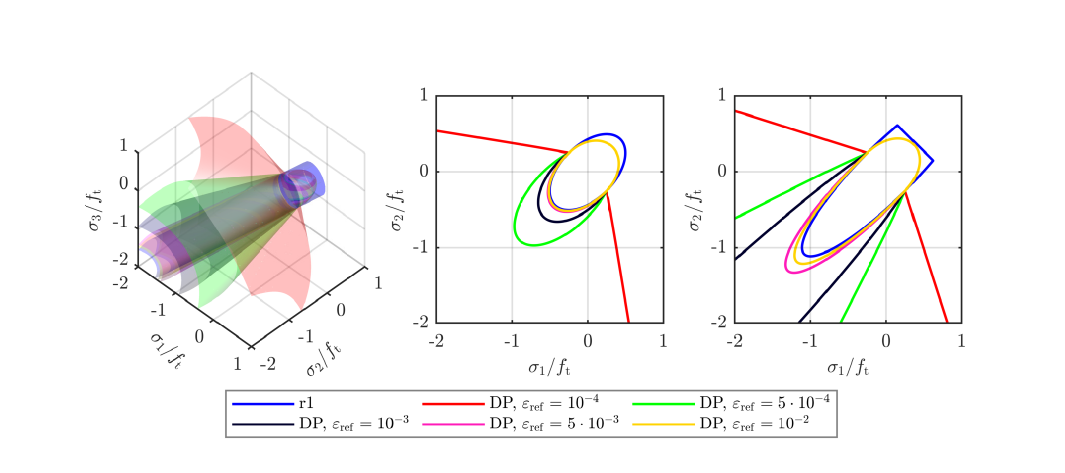}\\
    \begin{subfigure}{0.3\textwidth}
        \caption{Principal stress space}
    \end{subfigure}
    \begin{subfigure}{0.3\textwidth}
        \caption{Plane-stress ($\sigma_3=0$)}
    \end{subfigure}
     \begin{subfigure}{0.3\textwidth}
        \caption{Plane-strain ($\varepsilon_3=0$)}
    \end{subfigure}
    \caption{Strength surfaces for a fully intact material ($\phi=0$) using $f_\text{s}=0.5f_\text{t}$, with the $r1$ criterion, \cref{eq:r1}, shown in blue, and the other colours indicating the Drucker-Prager-like criteria, \cref{eq:DP}, using different values for $\varepsilon_\text{ref}$.}
    \label{fig:failureSurfs}
\end{figure}

Here, we will consider two combinations of strength potential and eigenstrain directions. The first is the $r1$ criterion used by Vicentini et al. \citep{Vicentini2026}, defined through the tensile strength $f_\text{t}$ and the shear strength $f_\text{s}$:
\begin{subequations} \label{eq:r1}
\begin{align}
    F_\text{d} &= f_\text{t} \left< \text{tr}(\m{\eta})\right>^+ + f_\text{s} \lVert \text{dev}(\m{\eta}) \rVert \qquad F_\text{i} = -10^6 f_\text{t} \left< \text{tr}(\m{\eta})\right>^- \\
    \m{G}_1 &= \frac{1}{3}\frac{\text{tr}(\m{\varepsilon})}{| \text{tr}(\m{\varepsilon}) |}\m{I} \qquad \m{G}_2 = \frac{\text{dev}(\m{\varepsilon})}{\lVert \text{dev}(\m{\varepsilon}) \rVert}
\end{align}
\end{subequations}
where the Macaulay brackets $\left< \square \right>^+$ and $\left< \square \right>^-$ are used to indicate that this term is only non-zero if the quantity is positive/negative respectively, thereby enforcing a different behaviour under compression/tension. Specifically, the compressive strength is set to $-10^6 \times$ the tensile strength, and this compressive strength is not degraded, thereby ensuring that even when the material is fully damaged, $d\approx 0$, it still correctly enforces a non-penetration condition on contact. Furthermore, this compressive strength is set to be negative, such that once projected in the compressive direction, it produces a positive value. The directions for the eigenstrains, $\m{G}_1$ and $\m{G}_2$, are used in a normalized form, such that $\lambda_i$ directly dictates the magnitude of these strains. This failure criterion is shown in \cref{fig:failureSurfs} as a blue line, with the compressive strength limit/non-penetration condition, resulting from $F_\text{i}$, not shown as this falls far outside the limits of the plot. 

The second strength potential used is a combination of the $r2$ criterion from \citep{Vicentini2026} under tension, and a Drucker-Prager-like cone under compression. This function is defined through the tensile and shear strengths and an additional parameter $\varepsilon_\text{ref}$ that dictates the strength increase under compression:
\begin{subequations} \label{eq:DP}
\begin{align}
    F_\text{d} &=\begin{cases} 
    \sqrt{f_\text{t}^2\text{tr}^2\left(\m{\eta}\right)+f_\text{s}^2\lVert \text{dev}\left(\m{\eta}\right) \rVert^2} \qquad \text{if} \qquad \text{tr}\left(\m{\varepsilon}\right)\geq0\\ 
    f_\text{s}\left(1-\frac{\text{tr}\left(\m{\varepsilon}\right)}{\varepsilon_\text{ref}}\right)\lVert \text{dev}\left(\m{\eta}\right) \rVert \qquad \text{if} \qquad \text{tr}\left(\m{\varepsilon}\right)<0\\ 
    \end{cases}\\
    \m{G} &= \begin{cases} 
    \frac{\m{\varepsilon}}{\lVert \m{\varepsilon} \rVert} \qquad \text{if} \qquad \text{tr}\left(\m{\varepsilon}\right)\geq0\\ 
    \frac{\text{dev}(\m{\varepsilon})}{\lVert \text{dev}(\m{\varepsilon}) \rVert} \qquad \text{if} \qquad \text{tr}\left(\m{\varepsilon}\right)<0\\ 
    \end{cases}
\end{align}
\end{subequations}
As we explicitly enforce that the eigenstrains are in the deviatoric direction under compression through $\m{G}$, we do not need to separately enforce non-penetration conditions. We also note the presence of the strains $\m{\varepsilon}$ within the definition of the strength potential, which is required to obtain the strengthening effect under compression. As $\text{tr}(\m{\eta})\geq0$ per definition, $\m{\eta}$ cannot be used to define the compressive strains acting on the material. This failure surface is shown in \cref{fig:failureSurfs} for different values of $\varepsilon_\text{ref}$.

\section{Computational implementation} \label{sec:implementation}
The governing equations, \cref{eq:momentum,eq:pf_evolution}, are solved using the finite element method, discretizing the displacement and phase-field variable as:
\begin{equation}
    \v{u} = \sum_\text{el}\m{N}_\text{u}\v{u}_\text{n} \qquad \phi = \sum_\text{el} \v{N}_\phi \v{\upphi}_\text{n}
\end{equation}
 using quadratic Lagrange elements for both the displacements, $\m{N}_\text{u}$, and the phase-field, $\v{N}_\phi$. $\v{u}_\text{n}$ and the phase-field $\v{\upphi}_\text{n}$ are the nodal values for the displacement and phase-field for the specific element being considered. For notational convenience of the derivatives of the strength surface, and the resulting tangent stiffness matrix, we will use Mandel notation, defining the relation between displacements and strains as:
\begin{equation}
    \v{\upvarepsilon} = \begin{bmatrix} \varepsilon_{xx} & \varepsilon_{yy} & \varepsilon_{zz} & \sqrt{2}\varepsilon_{yz} & \sqrt{2}\varepsilon_{xz} & \sqrt{2}\varepsilon_{xy}\end{bmatrix}^\top = \m{L}\m{N}_\text{u}\v{u}_\text{n}
\end{equation}
where $\m{L}$ is the displacement-to-strain mapping matrix. Using this discretization, the weak forms of the governing equation are discretized as:
\begin{equation} \label{eq:discretized_u}
    \v{F}_\text{u} = \int_\Omega \m{N}^\top_\text{u}\m{L}^\top\v{\upsigma} + \rho c \m{N}^\top_\text{u} \m{N}_\text{u} \v{\dot{u}}_\text{n} + \rho \m{N}^\top_\text{u} \m{N}_\text{u} \v{\ddot{u}}_\text{n} + \m{N}_\text{u}^\top \v{b}\;\text{d}\Omega + \int_\Gamma \m{N}_\text{u}^\top \v{\tau} \; \text{d}\Gamma = \v{0}
\end{equation}
\begin{equation}  \label{eq:discretized_phi}
    \v{F}_\phi = \int_\Omega \frac{G_\text{c}}{\ell}\v{N}_\phi^\top \v{N}_\phi \v{\upphi}_\text{n} + G_\text{c}\ell \left(\nabla \v{N}_\phi\right)^\top \nabla \v{N}_\phi \v{\upphi}_\text{n} - 2(1-\kappa)\v{N}_\phi^\top (1-\v{N}_\phi^\top \v{\upphi}_\text{n}) \mathcal{F}\;\text{d}\Omega = \v{0}
\end{equation}
where the stresses $\v{\upsigma}$ (and the tangent stiffness matrix required for the non-linear solver) and the crack driving force $\mathcal{F}$ are calculated within each integration point through a return-mapping scheme, similarly as is standard within plasticity models. A multi-pass staggered solution scheme is used, solving for the displacements and fracture eigenstrains first, then solving for the phase-field, and iterating several times until both fields are converged (or a maximum of 5 passes have been performed). A Newmark scheme is used for the time discretization of the displacements, using time-integration parameters $\beta=0.5625$ and $\gamma=1.0$. These values for the time-integration are chosen to enforce sufficient numerical damping, preventing stress waves travelling through the complete domain and reflecting on the boundaries. The discretized equations are implemented using the finite-element programming library FEniCSx \citep{Baratta2023}, together with the numba library \citep{Numba} for the calculation of per-integration-point quantities, and the codes used to generate all results are openly available (see data availability at the end of this paper). 

\subsection{Return mapping scheme}
The strength criterion, \cref{eq:StrengthCriteria}, together with the relation between stresses and elastic energy, \cref{eq:stresses}, are solved within each integration point. First considering the case with a single eigenstrain direction, e.g. \cref{eq:DP}, this scheme is described through the two residuals:
\begin{subequations} \label{eq:ReturnMapping_1Lambda}
\begin{align} 
    f_1 &=0= \v{\upsigma} - \frac{\partial \Psi_\text{el}}{\partial \v{\upvarepsilon}} = 0\\
    f_2 &=0= \begin{cases} 
    \lambda \qquad &\text{if} \qquad \left(\frac{\partial \v{\upeta}}{\partial \lambda}\right)^\top \left(\frac{\partial F}{\partial \v{\upeta}}-\v{\upsigma} \right) > 0\\
    \left(\frac{\partial \v{\upeta}}{\partial \lambda}\right)^\top \left( \frac{\partial F}{\partial \v{\upeta}}-\v{\upsigma}\right) + \kappa_\text{t}K \lambda \qquad &\text{otherwise}
    \end{cases}
\end{align}
\end{subequations}
where the $f_2=\lambda$ condition is used when the stresses are within the failure criterion, enforcing $\lambda=0$, while the second criterion solves for $\lambda$ such that the stresses are on the fracture criterion. A small additional factor $\kappa_\text{t}K\lambda$ is added to the second equation, with this term providing a residual stiffness post-fracture by extending the failure surface based on the magnitude of the fracture eigenstrains. This stiffness is needed in addition to the residual strength $\kappa$ from \cref{eq:d_e} to ensure no unconstrained degrees of freedom arise when the stresses are on the strength surface.

The system of equations from \cref{eq:ReturnMapping_1Lambda} is solved using an iterative Newton-Raphson scheme, described in the case where the stresses exceed the failure function by:
\begin{equation} 
    \m{K}\begin{bmatrix} \v{\text{d}\upsigma} \\ \text{d}\lambda \end{bmatrix} = -\begin{bmatrix} f_1 \\ f_2 \end{bmatrix} \qquad \text{where} \qquad \m{K} = \begin{bmatrix} \m{I} & \frac{\partial^2 \Psi}{\partial \v{\upvarepsilon}^2}\frac{\partial \v{\upeta}}{\partial \lambda}\\
    \left(\frac{\partial \v{\upeta}}{\partial \lambda}\right)^\top & \kappa_\text{t}K
    \end{bmatrix}
\end{equation}
using the tangent matrix $\m{K}$. This matrix is also used in constructing a consistent tangent stiffness matrix as:
\begin{equation}
    \begin{bmatrix} \frac{\partial \v{\upsigma}}{\partial \v{\varepsilon}} \\ \frac{\partial \lambda}{\partial \varepsilon} \end{bmatrix} = \m{K}^{-1} \begin{bmatrix} -\frac{\partial^2 \Psi}{\partial \v{\upvarepsilon}^2} \\ 0 \end{bmatrix}
\end{equation}
where the top $(6\times6)$ sub-matrix, corresponding to the $\frac{\partial \v{\upsigma}}{\partial \v{\varepsilon}}$, is used within the non-linear solving scheme for the momentum balance, \cref{eq:discretized_u}. For this case, where only a single direction of the fracture eigenstrains is used, this tangent stiffness matrix can also be explicitly expressed in terms of the direction of the eigenstrains as:
\begin{equation}
    \m{D}_\text{tan} =  \frac{\partial^2 \Psi_\text{el}}{\partial \v{\upvarepsilon}^2}  \left(\m{I} - \frac{\frac{\partial^2 \Psi_\text{el}}{\partial \v{\upvarepsilon}^2}\frac{\partial \v{\upeta}}{\partial \lambda} \left(\frac{\partial \v{\upeta}}{\partial \lambda}\right)^\top }{\left(\frac{\partial \v{\upeta}}{\partial \lambda}\right)\frac{\partial^2 \Psi_\text{el}}{\partial \v{\upvarepsilon}^2}\frac{\partial \v{\upeta}}{\partial \lambda} + \kappa_\text{t}K}\right) 
\end{equation}
which clearly shows the role of the residual stiffness $\kappa_\text{t}$. Without this term, as soon as the stresses exceed the strength surface the stiffness in the direction of the eigenstrains becomes zero. By including this term, the stiffness is set to a small value, avoiding issues with ill-constrained degrees of freedom. Furthermore, this slight offset from the strength surface also helps the iterative scheme from \cref{eq:ReturnMapping_1Lambda} by preventing the state from alternating between two conditions due to trying to return to the surface exactly. 

If instead we consider a non-smooth fracture criterion, e.g. \cref{eq:r1}, where the strength criterion is checked and the eigenstrain develops in multiple independent directions, the residuals are defined in terms of stresses $\v{\upsigma}$, and the two eigenstrain magnitudes $\lambda_1$ and $\lambda_2$ as:
\begin{subequations} \label{eq:ReturnMapping_2Lambda}
\begin{align} 
    f_1 &=0= \v{\upsigma} - \frac{\partial \Psi_\text{el}}{\partial \v{\upvarepsilon}} = 0\\
    f_2 &=0= \begin{cases} 
    \lambda_1 \qquad &\text{if} \qquad \left(\frac{\partial \v{\upeta}}{\partial \lambda_1}\right)^\top \left(\frac{\partial F}{\partial \v{\upeta}}-\v{\upsigma} \right) > 0\\
    \left(\frac{\partial \v{\upeta}}{\partial \lambda_1}\right)^\top \left( \frac{\partial F}{\partial \v{\upeta}}-\v{\upsigma}\right) + \kappa_\text{t}K \lambda_1 \qquad &\text{otherwise}
    \end{cases}\\
    f_3 &=0= \begin{cases} 
    \lambda_2 \qquad &\text{if} \qquad \left(\frac{\partial \v{\upeta}}{\partial \lambda_2}\right)^\top \left(\frac{\partial F}{\partial \v{\upeta}}-\v{\upsigma} \right) > 0\\
    \left(\frac{\partial \v{\upeta}}{\partial \lambda_2}\right)^\top \left( \frac{\partial F}{\partial \v{\upeta}}-\v{\upsigma}\right) + \kappa_\text{t}K \lambda_2 \qquad &\text{otherwise}
    \end{cases}
\end{align}
\end{subequations}
where for each direction the stresses need to stay within the individual strength surfaces, in this case separately checking for the volumetric and shear components. Following from these residuals, assuming eigenstrains develop in both directions the iterative solution scheme is given as:
\begin{equation} 
    \m{K}\begin{bmatrix} \v{\text{d}\upsigma} \\ \text{d}\lambda_1 \\ \text{d}\lambda_2 \end{bmatrix} = -\begin{bmatrix} f_1 \\ f_2 \\ f_3 \end{bmatrix} \qquad \text{where} \qquad \m{K} = \begin{bmatrix} \m{I} & \frac{\partial^2 \Psi}{\partial \v{\upvarepsilon}^2}\frac{\partial \v{\upeta}}{\partial \lambda_1} & \frac{\partial^2 \Psi}{\partial \v{\upvarepsilon}^2}\frac{\partial \v{\upeta}}{\partial \lambda_2}\\
    \left(\frac{\partial \v{\upeta}}{\partial \lambda_1}\right)^\top & \kappa_\text{t}K & 0 \\
    \left(\frac{\partial \v{\upeta}}{\partial \lambda_2}\right)^\top & 0 & \kappa_\text{t}K \\
    \end{bmatrix}
\end{equation}
If instead, only one of the failure criteria is exceeded, the tangent matrix becomes:
\begin{equation} 
    \m{K} = \begin{bmatrix} \m{I} & \frac{\partial^2 \Psi}{\partial \v{\upvarepsilon}^2}\frac{\partial \v{\upeta}}{\partial \lambda_1} & \frac{\partial^2 \Psi}{\partial \v{\upvarepsilon}^2}\frac{\partial \v{\upeta}}{\partial \lambda_2}\\
    \left(\frac{\partial \v{\upeta}}{\partial \lambda_1}\right)^\top & \kappa_\text{t}K & 0 \\
    0 & 0 & 1 \\
    \end{bmatrix} \qquad \text{or} \qquad
    \m{K} = \begin{bmatrix} \m{I} & \frac{\partial^2 \Psi}{\partial \v{\upvarepsilon}^2}\frac{\partial \v{\upeta}}{\partial \lambda_1} & \frac{\partial^2 \Psi}{\partial \v{\upvarepsilon}^2}\frac{\partial \v{\upeta}}{\partial \lambda_2}\\
    0 & 1 & 0 \\
    \left(\frac{\partial \v{\upeta}}{\partial \lambda_2}\right)^\top & 0 & \kappa_\text{t}K \\
    \end{bmatrix}
\end{equation}
when eigenstrains only develop in the $\lambda_1$ or $\lambda_2$ directions respectively. An explicit expression for the tangent stiffness matrix follows from using the Schur complement:
\begin{equation}
    \m{D}_\text{tan} =  \frac{\partial^2 \Psi_\text{el}}{\partial \v{\upvarepsilon}^2} \left(\m{I} -\frac{\partial^2 \Psi_\text{el}}{\partial \v{\upvarepsilon}^2}\m{g}\m{S}^{-1}\v{g}^\top \right) \qquad \text{where} \qquad \v{g} = \begin{bmatrix} \frac{\partial \v{\upeta}}{\partial \lambda_1} &  \frac{\partial \v{\upeta}}{\partial \lambda_2} \end{bmatrix} \qquad \text{and}\qquad \m{S} = \v{g}^\top \frac{\partial^2 \Psi_\text{el}}{\partial \v{\upvarepsilon}^2} \v{g} - \begin{bmatrix} \kappa_\text{t} K & 0 \\ 0 & \kappa_\text{t} K \end{bmatrix}
\end{equation}
If only the strength in one of the eigenstrain directions is exceeded, the direction that corresponds to the intact strength in $\v{g}$ is set to zero and the diagonal term is set to one to obtain the tangent stiffness. We note that the use of a Schur complement, rather than directly calculating the inverse of the $\m{K}$ matrix is needed due to the poor conditioning of this matrix. Convergence issues were encountered using $\m{K}^{-1}$, whereas no such issues were seen using the Schur complement which obtained well-converging results for all cases considered, with convergence rates between linear at the onset of fractures and quadratic when no new integration points start to fracture.

\section{Benchmark cases} \label{sec:results}
\begin{figure}
    \centering
    \begin{subfigure}{0.3\textwidth}
        \hspace{-0.5cm}
        \includegraphics{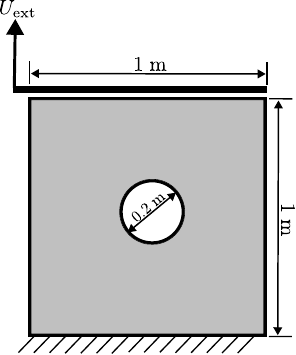}
        \caption{Plate with hole}
    \end{subfigure}
    \begin{subfigure}{0.3\textwidth}
        \hspace{-0.5cm}
        \includegraphics{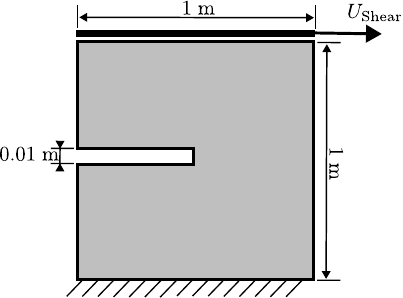}
        \caption{Single edge notched plate}
    \end{subfigure}
    \begin{subfigure}{0.3\textwidth}
        \hspace{0.2cm}
        \includegraphics{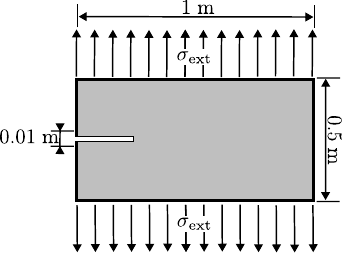}
        \caption{Dynamic fractures}
    \end{subfigure}
    \caption{Geometries and boundary conditions considered for the example cases.}
    \label{fig:geometries}
\end{figure}

\begin{table}[]
    \centering
    \begin{tabular}{| c c | c c |}
    \hline
        Young's modulus & $E$ & $200$ & $\text{GPa}$ \\
        Poisson ratio & $\nu$ & $0.3$ & \\
        Tensile strength & $f_\text{t}$ & 150 & $\text{MPa}$ \\
        Shear strength & $f_\text{s}$ & 150 & $\text{MPa}$ \\
        Fracture release energy & $G_\text{c}$ & $100$ & $\text{kJ}/\text{m}^2$\\
        Phase-field length scale & $\ell$ & $0.05$ & $\text{m}$\\
        Density & $\rho$ & $8000$ & $\text{kg}/\text{m}^3$\\
        Damping coefficient & $c$ & $10^{6}$ & $\text{s}^{-1}$\\
        Residual strength & $\kappa$ & $10^{-3}$ & \\
        Residual stiffness & $\kappa_\text{t}$ & $10^{-9}$ & \\
        \hline
    \end{tabular}
    \caption{Material properties used unless otherwise stated for \cref{sec:PlateHole,sec:SENT,sec:dynamic}.}
    \label{tab:matProps}
\end{table}

To demonstrate the capabilities of the presented model, and explore the required element sizes, and influence of model parameters, we will consider the three cases shown in \cref{fig:geometries}: A plate with a hole under tension, a notched specimen under shear, and a notched specimen under dynamic loading. All cases use the same material parameters, shown in \cref{tab:matProps}, unless stated otherwise, and all cases assume plane-strain conditions. 

\subsection{Plate with hole} \label{sec:PlateHole}
The first case considered consists of a unit square with a hole with a diameter of $0.2\;\text{m}$ in the centre. The bottom of the domain is fixed in horizontal and vertical direction, while a displacement is imposed at the top surface at a rate of $\dot{U}_\text{ext}=\pm3\cdot10^{-6}\;\text{m}/\text{s}$. This strain rate is chosen low to limit the impact of inertia for most cases, with it only affecting the obtained results for cases showing a snap-back type behaviour where the material suddenly fails. Loading under either tension (upwards displacement) or compression (downwards displacement) are considered. Due to the stress concentrations at the side of the hole, and the equal tensile and shear strength, cracks should form for both cases at the sides of the hole. For the tensile cases, these cracks will propagate horizontally, whereas for the compression cases they will follow the direction of the maximum shear stress, propagating at an approximate $45^\circ$ angle. 

\begin{figure}
    \centering
    \includegraphics{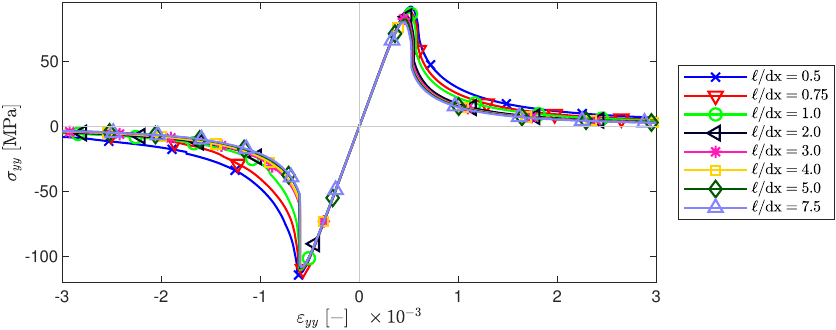}
    \caption{Effect of element size on load-displacement behaviour under compression (negative $\varepsilon_{yy}$) and tension (positive $\varepsilon_{yy}$).}
    \label{fig:platehole_Meshref}
\end{figure}

\begin{figure}
    \centering
    \begin{subfigure}{0.3\textwidth}
        \centering
        \includegraphics[clip=true,trim={10 10 20 0}]{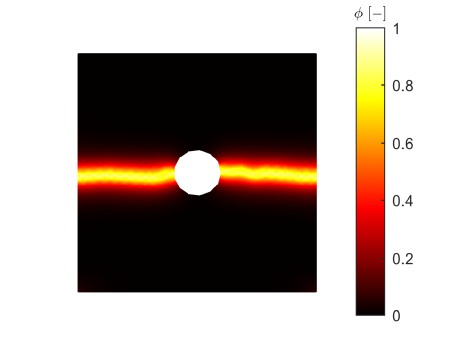}
        \caption{$\ell/\text{dx}=1$}
    \end{subfigure}
    \begin{subfigure}{0.3\textwidth}
        \centering
        \includegraphics[clip=true,trim={10 10 20 0}]{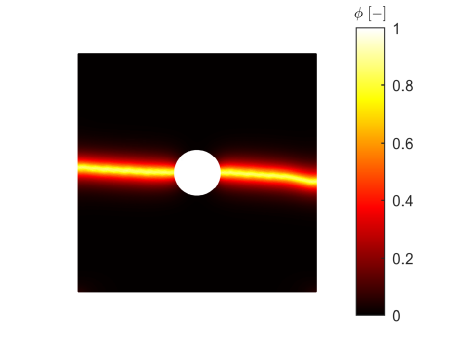}
        \caption{$\ell/\text{dx}=3$}
    \end{subfigure}
    \begin{subfigure}{0.35\textwidth}
        \centering
        \includegraphics[clip=true,trim={10 10 0 0}]{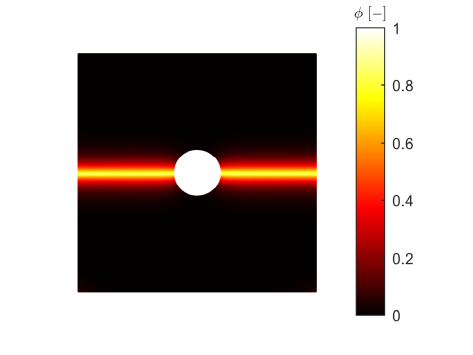}
        \caption{$\ell/\text{dx}=5$}
    \end{subfigure}
    \caption{Effect of element size on phase field for the tensile plate with hole case. Results shown at an applied strain of $\varepsilon_{yy}=3\cdot10^{-3}$.}
    \label{fig:PlateHole_dx_Tension}
\end{figure}

\begin{figure}
    \centering
    \begin{subfigure}{0.35\textwidth}
        \centering
        \includegraphics[scale=0.95,clip=true,trim={0 0 0 0}]{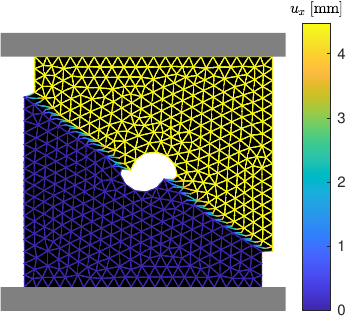}
        \caption{$\ell/\text{dx}=1$}
    \end{subfigure}
    \begin{subfigure}{0.35\textwidth}
        \centering
        \includegraphics[scale=0.95,clip=true,trim={0 0 0 0}]{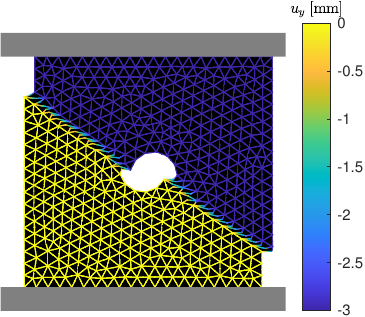}
        \caption{$\ell/\text{dx}=1$}
    \end{subfigure}
    \begin{subfigure}{0.25\textwidth}
        \centering
        \includegraphics[scale=0.95,clip=true,trim={25 15 0 0}]{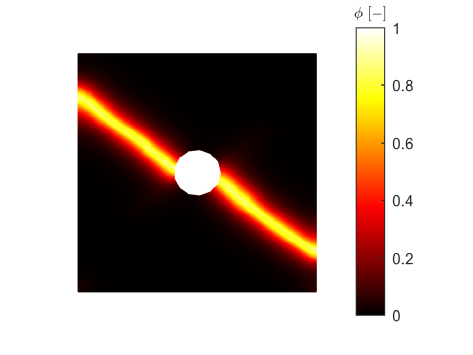}
        \caption{$\ell/\text{dx}=1$}
    \end{subfigure}
        \begin{subfigure}{0.35\textwidth}
        \centering
        \includegraphics[scale=0.95,clip=true,trim={0 0 0 0}]{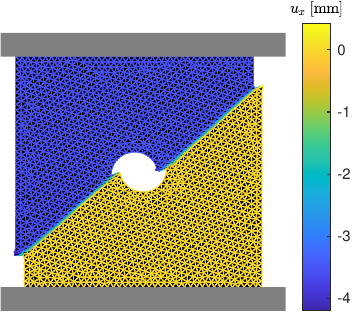}
        \caption{$\ell/\text{dx}=2$}
    \end{subfigure}
    \begin{subfigure}{0.35\textwidth}
        \centering
        \includegraphics[scale=0.95,clip=true,trim={0 0 0 0}]{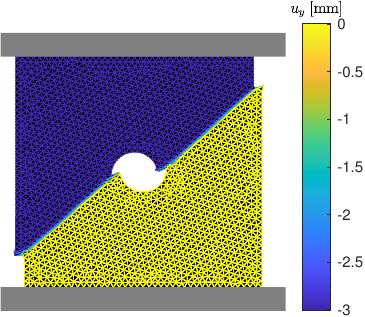}
        \caption{$\ell/\text{dx}=2$}
    \end{subfigure}
    \begin{subfigure}{0.25\textwidth}
        \centering
        \includegraphics[scale=0.95,clip=true,trim={25 15 0 0}]{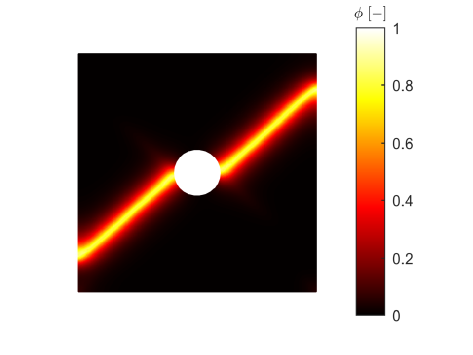}
        \caption{$\ell/\text{dx}=2$}
    \end{subfigure}
    \caption{Horizontal (a,d) and vertical (b,e) displacement, with deformations magnified by $\times10$ and showing the elements, and phase-field variable (c,f) for the compressive plate-with-hole case, using a coarse and a finer mesh.}
    \label{fig:PlateHole_dx_Compression}
\end{figure}

\subsubsection{Element-size requirements}
The guidance when using standard phase-field methods is to use $2-5$ elements per phase-field length scale. To confirm this still holds for the presented method, compression and tension simulations are performed for a range of element sizes, shown in \cref{fig:platehole_Meshref}. When the element size is larger than the phase-field length scale, this element size dictates the region over which the fracture energy is distributed, causing a more ductile behaviour. Starting from $\ell/\text{dx}\approx 2$, similar behaviour is observed for both tension and compression, with further mesh refinement having an almost negligible effect, especially for the compressive cases.

The phase-field obtained for the extensional case, shown in \cref{fig:PlateHole_dx_Tension} confirm this mesh dependence: When using $\ell/\text{dx}=1$ a clear effect of element orientation and size is observed, with the phase-field following the orientation of the elements, causing a sawtooth-like crack. Upon mesh refinement, this pattern disappears and a smooth phase-field is obtained starting from $\ell/\text{dx}=3$. However, the crack still shows a minor influence of the element orientation, observed by the crack not being horizontal, up to $\ell/\text{dx}=5$ where a horizontal crack is obtained. 

If we consider the compressive cases, \cref{fig:PlateHole_dx_Compression}, this element-orientation dependence is not exhibited as the crack path matches the element orientation, with $\ell/\text{dx}=2$ being sufficient to obtain mesh-independent results. Notably, displacement jump shows a clear localization to a single row of elements, and hence the thickness of the band over which the displacement jump is distributed becomes smaller with mesh refinement. However, the use of the phase-field distribution function to distribute the fracture release energy over a finite area suffices to regularize the problem and allow mesh-independent load-displacement behaviour to be obtained despite the displacement jump remaining localized. This case also shows that, for compressive cases, the fracture criterion correctly enforces a no-penetration condition, with the upper half of the domain slipping along the crack surface. 

\begin{figure}
    \centering
    \includegraphics{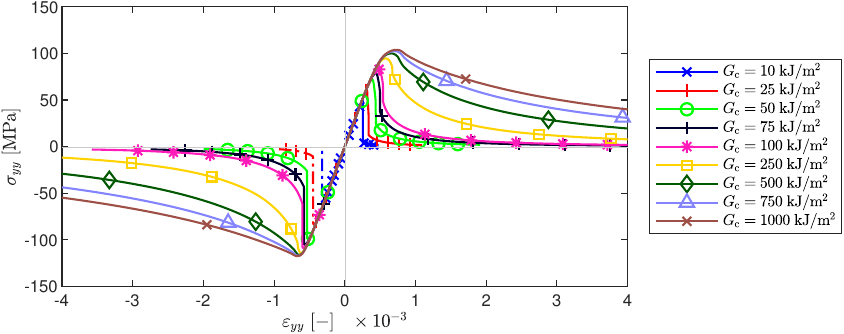}
    \caption{Effect of the fracture release energy $G_\text{c}$ on the load-displacement behaviour for the tensile and compressive plate with hole cases. Cases using dashed lines correspond to the horizontal failure mode discussed in the text.}
    \label{fig:PlateHole_Gc}
\end{figure}

\begin{figure}
    \centering
    \begin{subfigure}{0.4\textwidth}
        \centering
        \includegraphics[scale=0.95,clip=true,trim={0 15 0 0}]{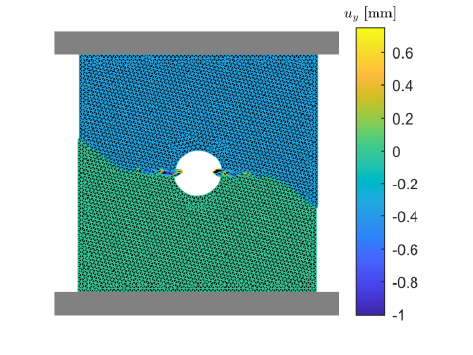}
        \caption{$G_\text{c}=10\;\text{kJ}/\text{m}^2$}
    \end{subfigure}
    \begin{subfigure}{0.4\textwidth}
        \centering
        \includegraphics[scale=0.95,clip=true,trim={0 15 0 0}]{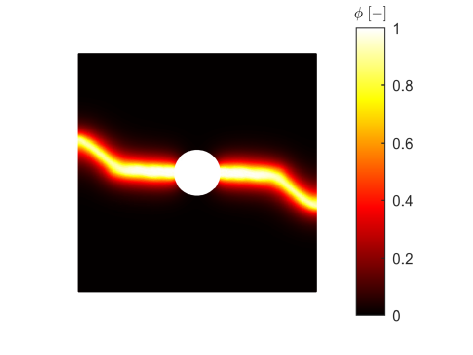}
        \caption{$G_\text{c}=10\;\text{kJ}/\text{m}^2$}
    \end{subfigure}
    \begin{subfigure}{0.4\textwidth}
        \centering
        \includegraphics[scale=0.95,clip=true,trim={0 15 0 0}]{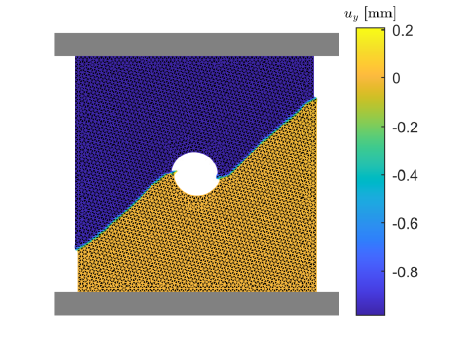}
        \caption{$G_\text{c}=25\;\text{kJ}/\text{m}^2$}
    \end{subfigure}
    \begin{subfigure}{0.4\textwidth}
        \centering
        \includegraphics[scale=0.95,clip=true,trim={0 15 0 0}]{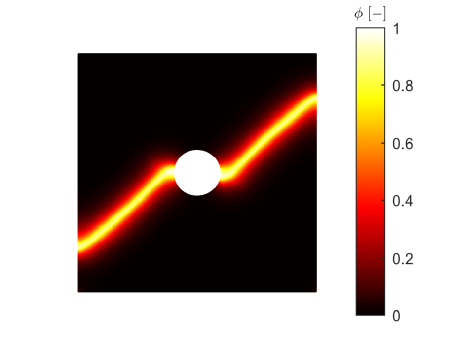}
        \caption{$G_\text{c}=25\;\text{kJ}/\text{m}^2$}
    \end{subfigure}
    \caption{Vertical displacement (a,c) with deformations magnified by $\times10$ and showing the elements, and phase-field variable (b,d) for the cases with low fracture release energy.}
    \label{fig:PlateHole_Gc_LocIssues}
\end{figure}

\subsubsection{Fracture release energy}
Next, we consider the effect of the fracture release energy, showing that the presented scheme correctly captures both the brittle behaviour expected at low energies, and the cohesive behaviour typical of higher fracture release energies. Simulations were performed for a range of energies between $G_\text{c} = 10\;\text{kJ}/\text{m}^2$ and $G_\text{c} = 1000\;\text{kJ}/\text{m}^2$, using a constant phase-field length scale of $\ell=0.05\;\text{m}$ and characteristic element size $\ell/\text{dx}=3$. 

The resulting load-displacement behaviour, shown in \cref{fig:PlateHole_Gc}, shows the clear impact of the fracture release energy. At low energies, an instantaneous drop in load is observed at the onset of fractures. As the energy is increased, this drop is reduced, and a more gradual unloading is observed, confirming that the model is able to correctly capture both the brittle and the cohesive regimes. For the lowest fracture energy under tension, where the behaviour is most brittle and the cohesive zone is negligible, the peak load can be compared to an analytical estimate. The stress concentration factor for a circular hole in an infinite plate under uniaxial tension is $K_\text{t}=3$ \citep{Timoshenko}, such that the expected applied stress at crack nucleation is $\sigma_\text{nuc} \approx f_\text{t}/K_\text{t} = 50\;\text{MPa}$. This is in agreement with the numerically obtained peak stress for the most brittle case in \cref{fig:PlateHole_Gc}, confirming that the prescribed strength surface correctly governs the onset of fracture. We further note that an increased peak load is observed for the higher values of $G_\text{c}$, which is a consequence of the cohesion: when the fracture energy is low, the material fails as soon as the strength surface is reached, causing a sharp crack to propagate. While the stresses are still bounded by the failure surface for the higher values of $G_\text{c}$, the cohesive zone behind the crack tip allows the total load transferred to increase as the crack propagates, while also reducing the stress concentration at the crack tip. As a result, the total load transferred increases initially as the crack propagates, causing the increased strength observed for higher values of $G_\text{c}$.

However, even though the load-displacement behaviour of all cases looks plausible, under compression the two cases with the lowest fracture release energy, $G_\text{c}=10\;\text{kJ}/\text{m}^2$ and $G_\text{c}=25\;\text{kJ}/\text{m}^2$, showed a different failure mode where the phase field did not localize into a diagonal shear band from the onset. Instead, cracks initially propagated horizontally with the phase-field spreading over a region roughly twice the width of properly localized cracks, as shown in \cref{fig:PlateHole_Gc_LocIssues}. The displacements obtained within these two cases also show a clearly different behaviour, where the material is crushed/squeezed horizontally, rather than slipping along a diagonal crack. As this horizontal damage zone propagates sideways, it eventually localizes into diagonal cracks. This behaviour was verified to be consistent across time-step refinements, phase-field length scales, staggered iteration counts, and both failure criteria considered, confirming that it is not a numerical artefact. A plausible explanation follows from the stress distribution around a circular hole under far-field compression \citep{Timoshenko}: a tensile hoop stress develops at the sides of the hole that decays rapidly with distance, alongside the far-field shear stress that sustains diagonal shear bands. At low $G_\text{c}$, the cohesive zone is negligible and the horizontal mode-I crack, driven by this tensile hoop stress, can propagate over a significant distance before the diminishing driving force arrests it. At higher $G_\text{c}$, the cohesive zone resistance arrests the horizontal crack after only short propagation, leaving the diagonal shear band, sustained by the far-field stress state, as the dominant mode. This competition between crack modes is only observed under compression, where both tensile hoop stresses and shear stresses coexist, and not under tension, where the stress state unambiguously drives a horizontal mode-I crack.

\begin{figure}
    \centering
    \begin{subfigure}{0.9\textwidth}
        \centering
        \includegraphics{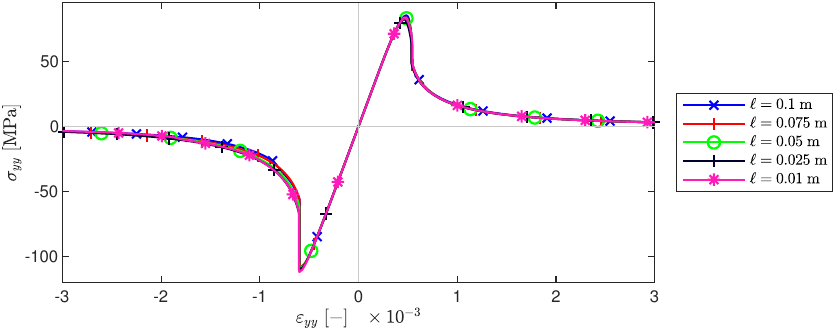}
        \caption{}
    \end{subfigure}
    \begin{subfigure}{0.45\textwidth}
        \centering
        \includegraphics{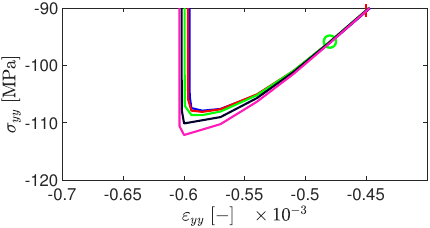}
        \caption{Compressive peak load}
    \end{subfigure}
    \begin{subfigure}{0.45\textwidth}
        \centering
        \includegraphics{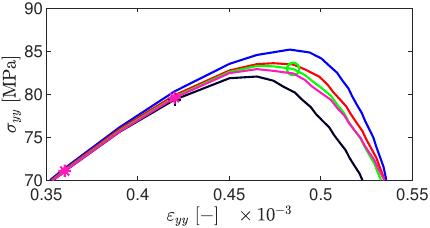}
        \caption{Tensile peak load}
    \end{subfigure}
    \caption{Effect of phase-field length scale on the load-displacement behaviour, and zooms on relevant regions.}
    \label{fig:ellRef}
\end{figure}

\begin{figure}
    \centering
    \begin{subfigure}{0.45\textwidth}
        \centering
        \includegraphics[scale=0.95,clip=true,trim={0 0 0 0}]{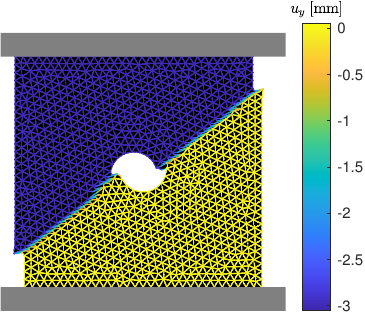}
        \caption{$\ell=0.1\;\text{m}$}
    \end{subfigure}
    \begin{subfigure}{0.45\textwidth}
        \centering
        \includegraphics[scale=0.95,clip=true,trim={0 15 0 0}]{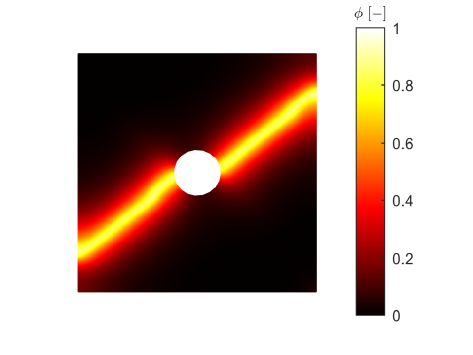}
        \caption{$\ell=0.1\;\text{m}$}
    \end{subfigure}
    \begin{subfigure}{0.45\textwidth}
        \centering
        \includegraphics[scale=0.95,clip=true,trim={0 10 0 0}]{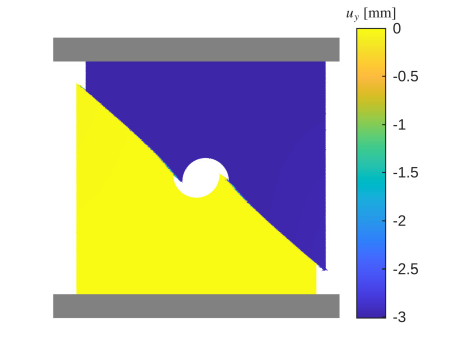}
        \caption{$\ell=0.01\;\text{m}$}
    \end{subfigure}
    \begin{subfigure}{0.45\textwidth}
        \centering
        \includegraphics[scale=0.95,clip=true,trim={0 15 0 0}]{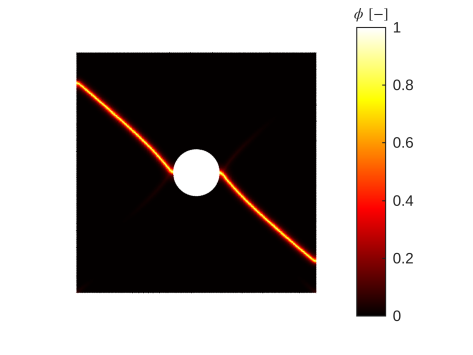}
        \caption{$\ell=0.01\;\text{m}$}
    \end{subfigure}
    \caption{Vertical displacements (a,c, deformations magnified by $\times10$) and phase-field (b,d) obtained for applied strains of $\varepsilon=3\cdot10^{-3}$, using different phase-field length scales.}
    \label{fig:ellRef_Phasefield}
\end{figure}

\subsubsection{Length-scale dependence}
Finally, we consider the effect of the phase-field length scale $\ell$, showing that within the presented model this is solely an indicator of the width of the phase-field region, rather than impacting the physical behaviour. Simulations were performed for a range of length scales between $\ell=0.01\;\text{m}$ and $\ell=0.1\;\text{m}$. These simulations were performed with a constant fracture release energy of $G_\text{c}=100\;\text{kJ}/\text{m}^2$ and characteristic element size $\ell/\text{dx}=3$. The resulting load displacement behaviour, \cref{fig:ellRef}, confirms that the phase-field length scale has negligible effect on the strength and energy dissipation during fracture. An order of magnitude difference in length scale (and element size) results in less than $5\%$ difference in the obtained peak load. While the unloading behaviour, especially for the compression cases, shows a slightly larger difference, with smaller length-scales experiencing a reduced drop in stress at the onset of fracture, this is still relatively minor and can be attributed to the reduced element size aiding with the localization of cracks and better resolving the increased stress concentrations. 

The obtained phase-field solutions and displacements, shown in \cref{fig:ellRef_Phasefield}, confirm that the length scale only impacts the width of the phase-field region. For both cases, the actual displacement discontinuity is limited to a single row of elements, despite the phase-field being distributed over a wider region for the larger length scales. Notably, even though the largest length scale is comparable to the size of the hole, it still localized properly to the sides of the hole, showing that as the phase-field is driven by the fracture-eigenstrains, it is able to localize to the correct regions even when the length scale is large. One notable effect of the length scale is on the computational cost, with the finest length scale requiring $100\times$ more elements than the coarsest length scale, and thus a significantly increased computational time (close to a day for $\ell=0.01\;\text{m}$ compared to around $5\;\text{min}$ for $\ell=0.1\;\text{m}$). This highlights the implications of having the length scale decoupled from the material strength: allowing for a length scale that is as large as possible to limit the computational cost of simulations, with the main restrictions following from the details of the features that are being resolved, while still obtaining the same crack paths and behaviour as would be obtained through fine length-scale simulations.

\begin{figure}
    \centering
    \includegraphics{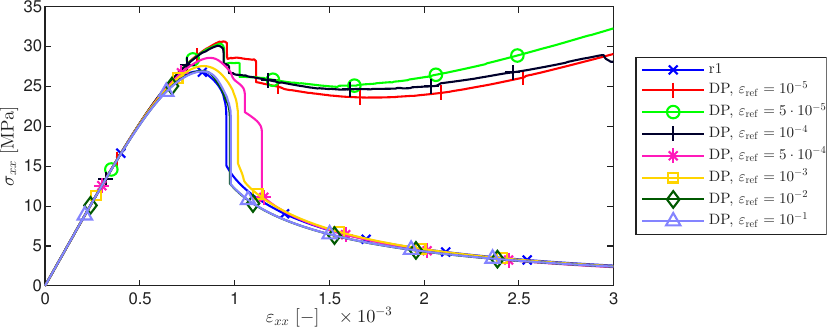}
    \caption{Load-displacement behaviour of the single-edge notched plate under shear loading, using the $r1$ and Drucker-Prager-like (referred to as ``DP'') strength criteria, for a range of reference strains.}
    \label{fig:SENT_LoadDisp}
\end{figure}

\begin{figure}
    \centering
    \begin{subfigure}{0.28\textwidth}
        \centering
        \includegraphics[scale=0.95,clip=true,trim={20 20 65 0}]{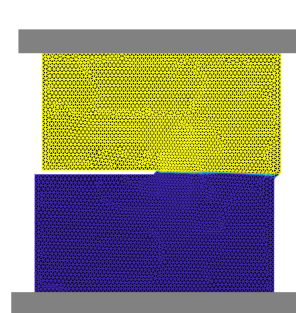}
        \caption{$r1$}
    \end{subfigure}
    \begin{subfigure}{0.28\textwidth}
        \centering
        \includegraphics[scale=0.95,clip=true,trim={20 20 65 0}]{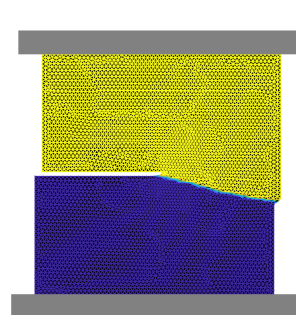}
        \caption{Drucker-Prager-like, $\varepsilon_\text{ref}=10^{-3}$}
    \end{subfigure}
    \begin{subfigure}{0.38\textwidth}
        \centering
        \includegraphics[scale=0.95,clip=true,trim={20 20 0 0}]{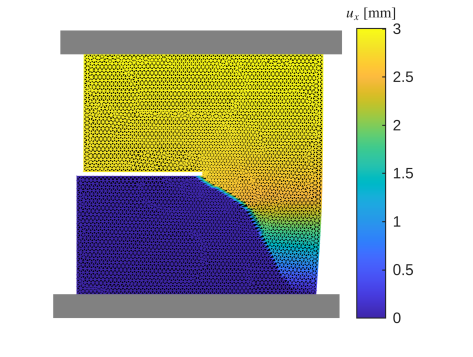}
        \caption{Drucker-Prager-like, $\varepsilon_\text{ref}=10^{-4}$}
    \end{subfigure}

    \begin{subfigure}{0.28\textwidth}
        \centering
        \includegraphics[scale=0.95,clip=true,trim={20 20 65 0}]{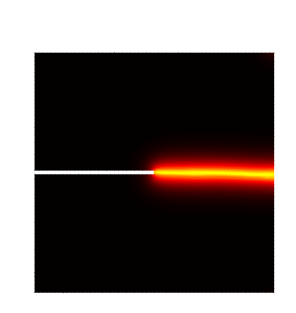}
        \caption{$r1$}
    \end{subfigure}
    \begin{subfigure}{0.28\textwidth}
        \centering
        \includegraphics[scale=0.95,clip=true,trim={20 20 65 0}]{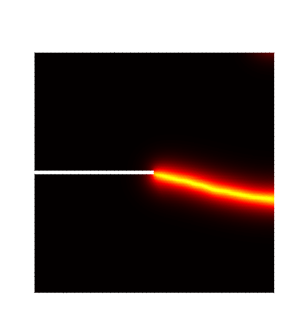}
        \caption{Drucker-Prager-like, $\varepsilon_\text{ref}=10^{-3}$}
    \end{subfigure}
    \begin{subfigure}{0.38\textwidth}
        \centering
        \includegraphics[scale=0.95,clip=true,trim={20 20 0 0}]{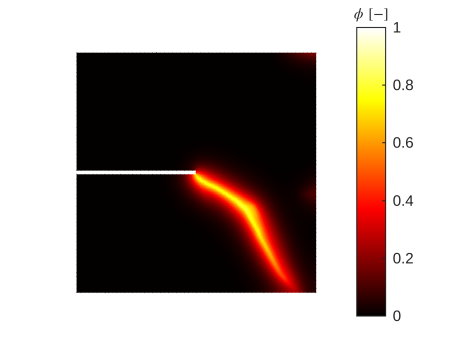}
        \caption{Drucker-Prager-like, $\varepsilon_\text{ref}=10^{-4}$}
    \end{subfigure}
    \caption{Horizontal displacements (a-c) and phase-field variable (d-f) obtained for the single-edge notched plate under shear loading at an applied strain of $\varepsilon_{xy}=3\cdot10^{-3}$, using the $r1$ and Drucker-Prager-like strength criteria.}
    \label{fig:SENT_results}
\end{figure}

\subsection{Single-edge notched plate} \label{sec:SENT}
Next, we consider a single-edge notched plate under shear loading, shown in \cref{fig:geometries}(b). The bottom of the domain is fixed in horizontal and vertical directions, the top surface is fixed in vertical direction while a horizontal displacement is imposed at a rate of $\dot{U}_\text{Shear}=1\cdot10^{-6}\;\text{m}/\text{s}$. Due to the stress concentration at the tip of the notch, a crack will form at this location. If fracture is driven by the shear stresses, the crack will propagate horizontally, while if it is driven by tensile stresses, the crack will propagate downwards under an approximate $45^\circ$ angle. Mixed-mode fracture will result in a fracture path in between these two extreme cases. We will consider both the $r1$ strength criterion from \cref{eq:r1} as well as the Drucker-Prager-like criterion from \cref{eq:DP} with varying reference strains $\varepsilon_\text{ref}$, controlling the increased strength in the compressive regime. 

The resulting load-displacement behaviour, shown in \cref{fig:SENT_LoadDisp}, indicates two distinct regimes for crack propagation: Using the $r1$ criterion or the Drucker-Prager-like criterion with a high reference strain, a sudden drop in load is observed during crack propagation. Representative results for this regime are shown in \cref{fig:SENT_results}(a,d), showing that the crack propagation is driven by shear. In contrast, using the Drucker-Prager-like criterion with a low reference strain, \cref{fig:SENT_results}(c,f), and thus a higher shear strength in compressive regions, the shear crack is partially suppressed with the initial crack propagating in a mixed-mode manner, before switching to a mode-I crack only. This competition between shear and tensile crack propagation is also observed in the load-displacement behaviour, showing small drops in strength during the mixed-mode propagation, while showing a hardening-like behaviour during later propagation, where the crack is driven by tensile stresses only. 

Notably, all cracks propagate either horizontally or downwards, with none of the cases showing cracks propagating upwards. This is a consequence of the crack driving force expressions: The r1 expression from \cref{eq:r1} explicitly removing compressive strains from the driving force by putting these in a non-damaged term, $F_\text{i}$, such that mode-I cracks are solely driven by extensional stresses. The resulting driving force in the compressive regime is solely due to deviatoric stresses, hence causing horizontal crack propagation. For the Drucker-Prager-like criterion, \cref{eq:DP}, volumetric strains do not contribute to crack propagation as they are not aligned with the eigenstrain directions under compressive loading via the direction $\v{G}$. While the approaches to eliminate the contribution of compressive stresses to crack propagation differ between the two criteria, they both successfully achieve this, resulting in no cracks propagating upwards.

\begin{figure}
    \centering
    \includegraphics{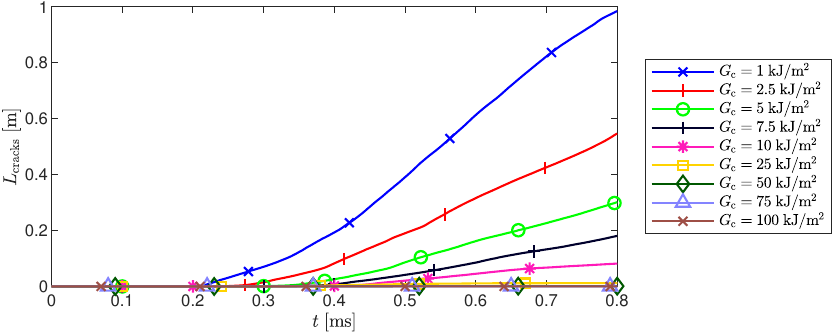}
    \caption{Impact of fracture energy release rate $G_\text{c}$ on the combined crack length for the dynamic fracture case.}
    \label{fig:Dynamic_Gc}
\end{figure}

\begin{figure}
    \centering
    \begin{subfigure}{0.45\textwidth}
        \centering
        \includegraphics[scale=1.0,clip=true,trim={0 50 0 35}]{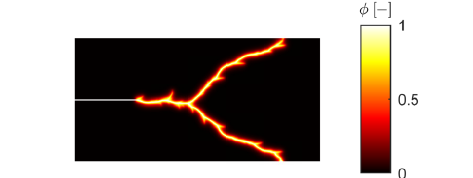}
        \caption{$G_\text{c}=1\;\text{kJ}/\text{m}^2$}
    \end{subfigure}
    \begin{subfigure}{0.45\textwidth}
        \centering
        \includegraphics[scale=1.0,clip=true,trim={0 50 0 35}]{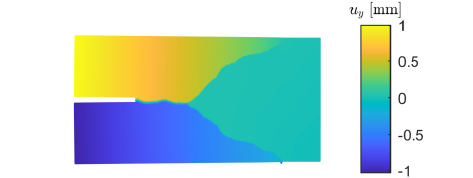}
        \caption{$G_\text{c}=1\;\text{kJ}/\text{m}^2$}
    \end{subfigure}
    \begin{subfigure}{0.45\textwidth}
        \centering
        \includegraphics[scale=1.0,clip=true,trim={0 50 0 35}]{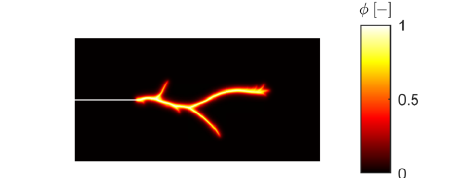}
        \caption{$G_\text{c}=2.5\;\text{kJ}/\text{m}^2$}
    \end{subfigure}
    \begin{subfigure}{0.45\textwidth}
        \centering
        \includegraphics[scale=1.0,clip=true,trim={0 50 0 35}]{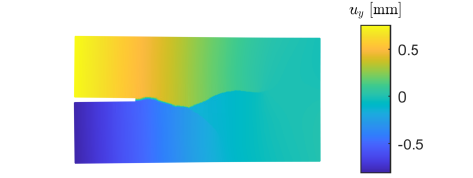}
        \caption{$G_\text{c}=2.5\;\text{kJ}/\text{m}^2$}
    \end{subfigure}
    \begin{subfigure}{0.45\textwidth}
        \centering
        \includegraphics[scale=1.0,clip=true,trim={0 50 0 35}]{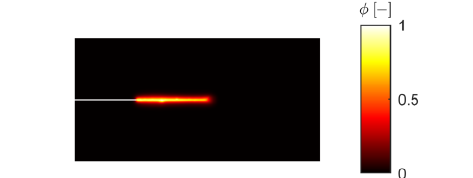}
        \caption{$G_\text{c}=7.5\;\text{kJ}/\text{m}^2$}
    \end{subfigure}
    \begin{subfigure}{0.45\textwidth}
        \centering
        \includegraphics[scale=1.0,clip=true,trim={0 50 0 35}]{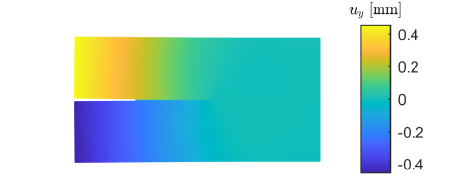}
        \caption{$G_\text{c}=7.5\;\text{kJ}/\text{m}^2$}
    \end{subfigure}
    \begin{subfigure}{0.45\textwidth}
        \centering
        \includegraphics[scale=1.0,clip=true,trim={0 50 0 35}]{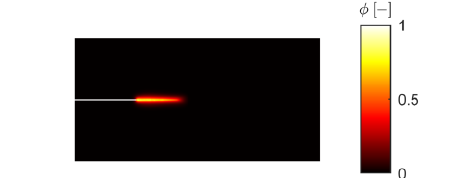}
        \caption{$G_\text{c}=10.0\;\text{kJ}/\text{m}^2$}
    \end{subfigure}
    \begin{subfigure}{0.45\textwidth}
        \centering
        \includegraphics[scale=1.0,clip=true,trim={0 50 0 35}]{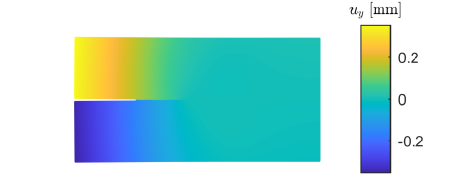}
        \caption{$G_\text{c}=10.0\;\text{kJ}/\text{m}^2$}
    \end{subfigure}
    \caption{Phase-field (left column) and vertical displacement (right column, magnified by $\times1000$) obtained for the dynamic fracture case at $t=8\;\text{ms}$.}
    \label{fig:dynamic_plots}
\end{figure}

\subsection{Dynamic fracture} \label{sec:dynamic}

The final case considered consists of a notched plate of $1\;\text{m}\times0.5\;\text{m}$ with an initial notch of $0.25\;\text{m}$ at the centre-left edge, shown in \cref{fig:geometries}(c). An external traction of $\sigma_\text{ext}=10\;\text{MPa}$ is applied to the top and bottom edges. As this case considers the dynamic behaviour, we set the damping coefficient to $c=0$. To give a detailed look into the crack paths, we also use a smaller phase-field length scale of $\ell=0.01\;\text{m}$, with a corresponding element size of $\text{dx}=\ell/3$. The fracture criterion uses the Drucker-Prager-like criterion from \cref{eq:DP} with a reference strain of $\varepsilon_\text{ref}=1$, practically removing the pressure dependence while retaining the smooth transition between mode-I and mode-II cracks. The fracture release energy $G_\text{c}$ is varied between $1\;\text{kJ}/\text{m}^2$ and $100\;\text{kJ}/\text{m}^2$, covering purely brittle crack propagation to a more cohesive behaviour. To get a measure of the length of the developed cracks, we estimate the total crack length through the phase-field distribution function $\gamma$, \cref{eq:pf_Gamma}, as:
\begin{equation}
    L_\text{crack} \approx \int_\Omega \gamma(\phi) \; \text{d}\Omega
\end{equation}
where we note that this only gives an estimate of the crack length for ductile cases. This is due to the phase-field distribution function being defined as the degradation of the fracture energy, rather than the presence of cracks as in standard phase-field methods. Hence, when the material has a noticeable cohesive zone, where $\phi<1$ along the crack, the obtained length will be an underestimate of the actual crack length. For brittle cases, where the phase-field is $\phi\approx1$ along the crack, this estimate is more accurate. 

The evolution of the crack length over time is shown in \cref{fig:Dynamic_Gc}. As expected, the cases with the lowest fracture release energy have the fastest crack propagation, reaching the edges of the domain after $8\;\text{ms}$. Increasing the value of the fracture energy reduces the rate of crack propagation, with the $G_\text{c}=10\;\text{kJ}/\text{m}^2$ case showing approximately $10\% $ of the crack length of the $G_\text{c}=1\;\text{kJ}/\text{m}^2$ case. The obtained fracture pattern for the $G_\text{c}=1\;\text{kJ}/\text{m}^2$ case is shown in \cref{fig:dynamic_plots}(a,b), showing the crack develops short branches as it propagates to dissipate additional energy, and branches into two distinct cracks. As the fracture release energy is increased, \cref{fig:dynamic_plots}(c,d), the formation of distinct branches is suppressed, with the branches that develop halting their propagation soon after forming. Further increases to the fracture release energy, \cref{fig:dynamic_plots}(e,f) and \cref{fig:dynamic_plots}(g,h) fully suppress all crack branching, resulting in a single crack propagating horizontally from the initial notch. As the fracture energy is increased beyond $G_\text{c}=10\;\text{kJ}/\text{m}^2$, the fracture becomes unable to propagate, instead creating a small region at the crack tip where the phase-field is increased. 
% ONLY VALID FOR STATIC CASES, NOT DYNAMIC ONES
%This threshold can be compared to the Griffith criterion for an edge crack in a semi-infinite plate under plane strain, which predicts a critical fracture energy of:
%\begin{equation}
%    G_\text{c,crit} = \frac{\sigma_\text{ext}^2 \pi a}{E'} = \frac{(10\;\text{MPa})^2 \pi (0.25\;\text{m})}{200\;\text{GPa}/(1-0.3^2)} \approx 0.43\;\text{kJ}/\text{m}^2
%\end{equation}
%using the plane-strain modulus $E'=E/(1-\nu^2)$, the applied traction $\sigma_\text{ext}=10\;\text{MPa}$, and the initial notch length $a=0.25\;\text{m}$. This value lies between the $G_\text{c}=10\;\text{kJ}/\text{m}^2$ case, where the crack still propagates, and the $G_\text{c}=25\;\text{kJ}/\text{m}^2$ case, where it does not. However, this analytical estimate assumes a static, steady-state configuration and does not account for the dynamic stress amplification that arises when the traction is applied suddenly, which increases the effective driving force and thus allows cracks to propagate at higher values of $G_\text{c}$ than the static prediction. Accounting for this dynamic amplification, the observed threshold is consistent with the analytical prediction. 
We note that all the cases use the same strength surface, using the same values for the tensile and shear strength, hence showing that these do not play a significant role in determining when cracks can propagate. This is in contrast to the previous cases that studied crack initialization, where the strength surface and tensile strength played a dominant role. As a result, this indicates that the presented scheme correctly captures the physical behaviour of crack nucleation and propagation, with the strength surface governing the former and the fracture release energy governing the latter, with both of these aspects being independent of the phase-field length scale used for the numerical regularization.

\section{Conclusions}

In this paper, we presented a computational framework for modelling cohesive fractures within the phase-field fracture paradigm. Using the concept of fracture eigenstrains from \citep{Vicentini2026} and integrating this within a local constitutive model, we are able to capture both brittle and cohesive fracture nucleation and propagation. The presented method is able to use a strength surface to predict the stress required for fracture nucleation, while the energy release rate governs the propagation behaviour. Furthermore, due to the used local formulation, no additional degrees of freedom are required to capture the cohesive behaviour compared to standard phase-field formulations, with the complete cohesive behaviour solved on a per-integration-point basis comparable to plasticity models. As a result, the presented model is straightforward to integrate within existing codes, and an example implementation is provided (see data availability statement).

The three benchmark cases confirm that crack nucleation is governed by the prescribed strength surface, while propagation is controlled by the fracture energy release rate $G_\text{c}$, with both aspects independent of the phase-field length scale $\ell$. Specifically, varying $\ell$ over an order of magnitude (from $0.01\;\text{m}$ to $0.1\;\text{m}$) produced less than $5\%$ variation in peak load for both tension and compression. The element-size study showed that mesh-independent load-displacement responses are obtained for $\ell/\text{dx}\geq 2$, consistent with the guidance for standard phase-field methods. For the dynamic branching case, reducing $G_\text{c}$ from $10\;\text{kJ}/\text{m}^2$ to $1\;\text{kJ}/\text{m}^2$ increased the crack length by approximately an order of magnitude and triggered crack branching, demonstrating the model's ability to capture complex fracture phenomena without ad-hoc criteria. The presented model thus provides a powerful and flexible tool for simulating cohesive fracture processes, with the potential for application across a wide range of materials and loading conditions.

\section*{Data availability}
\noindent The finite element code used within this study, together with all input files required to reproduce the results presented in this paper, is available at \url{https://github.com/T-Hageman/Cohesive-PhaseField-FenicsX}.

\section*{Acknowledgements}
\noindent No external funding was received for this research.

\section*{Declaration of competing interest}
\noindent The author declares that they have no known competing financial interests or personal relationships that could have appeared to influence the work reported in this paper.

\section*{CRediT authorship contribution statement}
\textbf{Tim Hageman}: Conceptualization, Methodology, Software, Validation, Formal analysis, Investigation, Data curation, Writing - original draft, Visualization.

\section*{Declaration of generative AI in the writing process}
\noindent During the preparation of this manuscript, the author used GitHub Copilot (Claude, Anthropic) to assist with improving the language and readability of the text, and provide critical feedback on draft versions. The author reviewed and edited all AI-generated suggestions and takes full responsibility for the content of the published work.

\bibliography{References}

\end{document}